\documentclass[a4paper,11pt]{article}

\pdfoutput=1

\usepackage{pos}
\usepackage{multicol}
\usepackage{wrapfig}
\usepackage{caption}
\usepackage{float,dsfont}
\usepackage{subcaption}

\newcommand{\Dslash}{{\not{\hspace{-0.1cm}D}}}
\newcommand{\zslash}{{\not{\hspace{-0.08cm}z}}} 
\def\MSbar{\overline{\rm MS}}
\def\slashed{{/}\mskip-10.0mu}
\def\openone{\leavevmode\hbox{\small1\kern-6.8pt\normalsize1}}

\title{Supercurrent Renormalization in $\cal{N}$ = 1 Supersymmetric Yang-Mills Theory}
\ShortTitle{Supercurrent renormalization}

\author[a]{G. Bergner}
\author[b]{M. Costa}
\author[b]{H. Panagopoulos}
\author[c]{S. Piemonte}
\author*[b]{A. Skouroupathis}
\author*[a]{I. Soler}
\author[b,d]{G. Spanoudes}

\affiliation[a]{University of Jena, Institute for Theoretical Physics,\\
Max-Wien-Platz 1, D-07743 Jena, Germany}

\affiliation[b]{Department of Physics, University of Cyprus,\\
1 Panepistimiou Avenue, 2109 Aglantzia, Nicosia, Cyprus}

\affiliation[c]{University of Regensburg, Institute for Theoretical Physics,\\
Universit\"{a}tsstr. 31, D-93040 Regensburg, Germany}

\affiliation[d]{Present Address: Computation-based Science
and Technology Research Center,\\
The Cyprus Institute, 20 Kavafi Str., Nicosia 2121, Cyprus}

\emailAdd{georg.bergner@uni-jena.de}
\emailAdd{kosta.marios@ucy.ac.cy}
\emailAdd{panagopoulos.haris@ucy.ac.cy}
\emailAdd{stefano.piemonte@ur.de}
\emailAdd{skouroupathis.apostolos@ucy.ac.cy}
\emailAdd{ivan.soler.calero@uni-jena.de}
\emailAdd{spanoudes.gregoris@ucy.ac.cy}

\DeclareMathOperator{\tr}{tr}

\abstract{In this work we study the renormalization of the SUSY Noether current in Supersymmetric $\cal{N}$ = 1 Yang-Mills (SYM) theory on the lattice. In particular, we study the mixing of the current with all other compatible operators of dimension 7/2 and 5/2, leading from the lattice-regularized to the $\MSbar$-renormalized operator basis. We perform our task in two ways: \\
(a) We compute, in dimensional regularization, the conversion factors relating the $\MSbar$ scheme to an intermediate gauge-invariant coordinate-space scheme. In this second scheme, renormalization can be performed via lattice simulations. This could help to investigate the breaking of SUSY on the lattice and strategies towards simulations of supersymmetric QCD. Here we present some preliminary numerical results.  \\
(b) We use lattice perturbation theory and compute, to one loop, various two- and three-point functions. We consider mixing with all relevant gauge-noninvariant operators, which contain also ghost fields.}

\FullConference{%
 The 38th International Symposium on Lattice Field Theory, LATTICE2021
  26th-30th July, 2021
  Zoom/Gather@Massachusetts Institute of Technology
}

\begin{document}
\maketitle

\section{Introduction}
The study of Supersymmetry on the Lattice is becoming more viable in recent years. One of the simplest models involving gauge fields, and an important forerunner for beyond-the-Standard-Model studies, is the Supersymmetric Yang-Mills (SYM) theory. A fundamental
question to be addressed is the breaking and restoration of SUSY on the lattice; to this effect, studies of the Noether supercurrent and its Green's functions are essential. Such studies necessitate a careful renormalization of the supercurrent and its complex pattern of mixing with other operators.

In this work we perform one-loop calculations of Green's functions (GF) containing the supercurrent, $S_\mu$, as well as other operators, which have the same or lower dimensionality with $S_\mu$ but they have the same quantum numbers. We employ dimensional regularization, in order to obtain the $\MSbar$-renormalized Green's functions, and lattice regularization for the extraction of the lattice renormalization functions in the $\MSbar$ scheme.  We use a standard discretization where gluinos live on the lattice sites and the gluons live on the link variables of the lattice. We obtain analytic expressions for the renormalization function of the supercurrent. The number of colors, $N_c$, and the gauge parameter, $\alpha$, are left unspecified.  Finally, we address the mixing which occurs among the operators and the supercurrent beyond tree level. In the same context, a gauge invariant renormalization scheme (GIRS)~\cite{Costa:2021iyv} is used to extract the renormalization of $S_\mu$; the renormalization factors and the mixing coefficients in GIRS scheme can be obtained also in a fully non-perturbative manner and the GIRS non-perturbative results can be connected with $\MSbar$ via conversion factors.

Our studies have been done in the Wess-Zumino (WZ) gauge. In this gauge, the SYM Lagrangian contains the gluon together with the gluino fields; the auxiliary fields are eliminated. The Lagrangian of SYM, in Euclidean space, is:
\begin{equation} 
\mathcal{L}_{\rm SYM}=\frac{1}{4}u_{\mu \nu}^{\alpha}u_{\mu \nu}^{\alpha}+\frac{1}{2}\bar{\lambda}^{\alpha}\gamma_{\mu}\mathcal{D}_{\mu}\lambda^{\alpha},
\label{susylagr}
\end{equation}

In order to perform our calculations, we fix the gauge by including a gauge-fixing term, together with the compensating ghost field $c^\alpha$ term; these terms are the same as in the non-supersymmetric case. The total action is no longer gauge invariant but it is Becchi-Rouet-Stora-Tyutin (BRST) invariant. The supercurrent is a spinor of dimension-$7/2$, which is gauge-invariant and carries one external Lorentz index, $\mu$:
\begin{equation}
S_\mu = -\frac{1}{2}{\rm tr}_c ( u_{\rho\,\sigma} [\gamma_\rho,\gamma_\sigma] \gamma_\mu \lambda )
\end{equation}

$S_\mu$ mixes with another dimension-7/2 gauge invariant operator $T_\mu$ \cite{Ali:2018fbq}:
\begin{equation}
T_\mu = 2 {\rm tr}_c(u_{\mu\,\nu} \gamma_\nu \lambda) 
\end{equation}

A consequence of gauge fixing is that $S_\mu$ can mix with three more classes of gauge-noninvariant operators which have the same transformation properties under global symmetries (e.g. Lorentz, or hypercubic on the lattice, global $SU(N_c)$ transformations, etc.) and whose dimension is lower or equal to that of $S_\mu$.
\begin{description}
\item [Class A:] BRST variations of some operator.
\item [Class B:] Operators which vanish by the equations of motion.
\item [Class C:] Any other operators which share the same global symmetries, but do not belong to the above classes; these can at most have finite mixing with $S_\mu$.
\end{description}
We present all candidate gauge-noninvariant operators which can mix with $S_\mu$ and belong to the classes A, B, C:
\bigskip
\begin{multicols}{2}
\begin{minipage}[t]\columnwidth
\vskip -1.5cm
\begin{eqnarray}
{\cal O}_{A1} &=&  \frac{1}{\alpha} {\rm tr}_c (\partial_{\nu} u_\nu \gamma_\mu \lambda) - ig  \, {\rm{tr}}_c( [c, \bar{c}]   \gamma_\mu \lambda) \nonumber \\
{\cal O}_{B1} &=&  {\rm tr}_c (u_\mu \Dslash \lambda)\nonumber \\
{\cal O}_{B2} &=&  {\rm tr}_c (\slashed{u} \gamma_\mu \Dslash \lambda)\nonumber \\
{\cal O}_{C1} &=&  {\rm tr}_c (u_\mu \lambda) \nonumber \\
{\cal O}_{C2} &=& {\rm tr}_c (\slashed{u}  \gamma_\mu \lambda) \nonumber \\
{\cal O}_{C3} &=&  {\rm tr}_c (\slashed{u} \partial_{\mu} \lambda) \nonumber
\end{eqnarray}
\end{minipage}
\begin{minipage}[t]\columnwidth
\vskip -1.5cm
\begin{eqnarray}
{\cal O}_{C4} &=&  {\rm tr}_c ((\partial_{\mu} \slashed{u}) \, \lambda) \nonumber \\
{\cal O}_{C5} &=&  {\rm tr}_c ((\partial_{\nu} u_\nu) \gamma_{\mu} \lambda) \nonumber  \\ 
{\cal O}_{C6} &=&  {\rm tr}_c (u_\nu \gamma_{\mu} \partial_{\nu} \lambda) \nonumber \\
{\cal O}_{C7} &=& i\, g \,{\rm tr}_c ([u_{\rho\,},u_{\sigma}] [\gamma_\rho,\gamma_\sigma] \gamma_\mu  \lambda) \nonumber \\
{\cal O}_{C8} &=& i\,g \,{\rm tr}_c ( [u_{\mu},u_{\nu}] \gamma_\nu \lambda ) \nonumber \\
{\cal O}_{C9} &=& i\,g\, {\rm tr}_c ([c, \bar{c}] \gamma_\mu \lambda)
\label{All_operators}
\end{eqnarray}
\end{minipage}
\end{multicols}

\section{Feynman diagrams}
The extraction of all mixing coefficients entering the renormalization of $S_\mu$ entails evaluation of the two- and three-point GFs: $\langle u_\nu S_\mu \bar \lambda  \rangle$, $\langle c \, S_\mu \, \bar c \, \bar \lambda \rangle$ and $\langle u_\nu u_\rho S_\mu \bar \lambda  \rangle$. As an example of the Feynman diagrams which must be computed at one loop, we show in Fig.~\ref{fig3ptguu} those relevant to $\langle u_\nu u_\rho S_\mu \bar \lambda  \rangle$. The renormalization of $T_\mu$ involves the same Green's functions, with $S_\mu$ replaced by $T_\mu$.

\begin{center}
\includegraphics[width=0.80\linewidth]{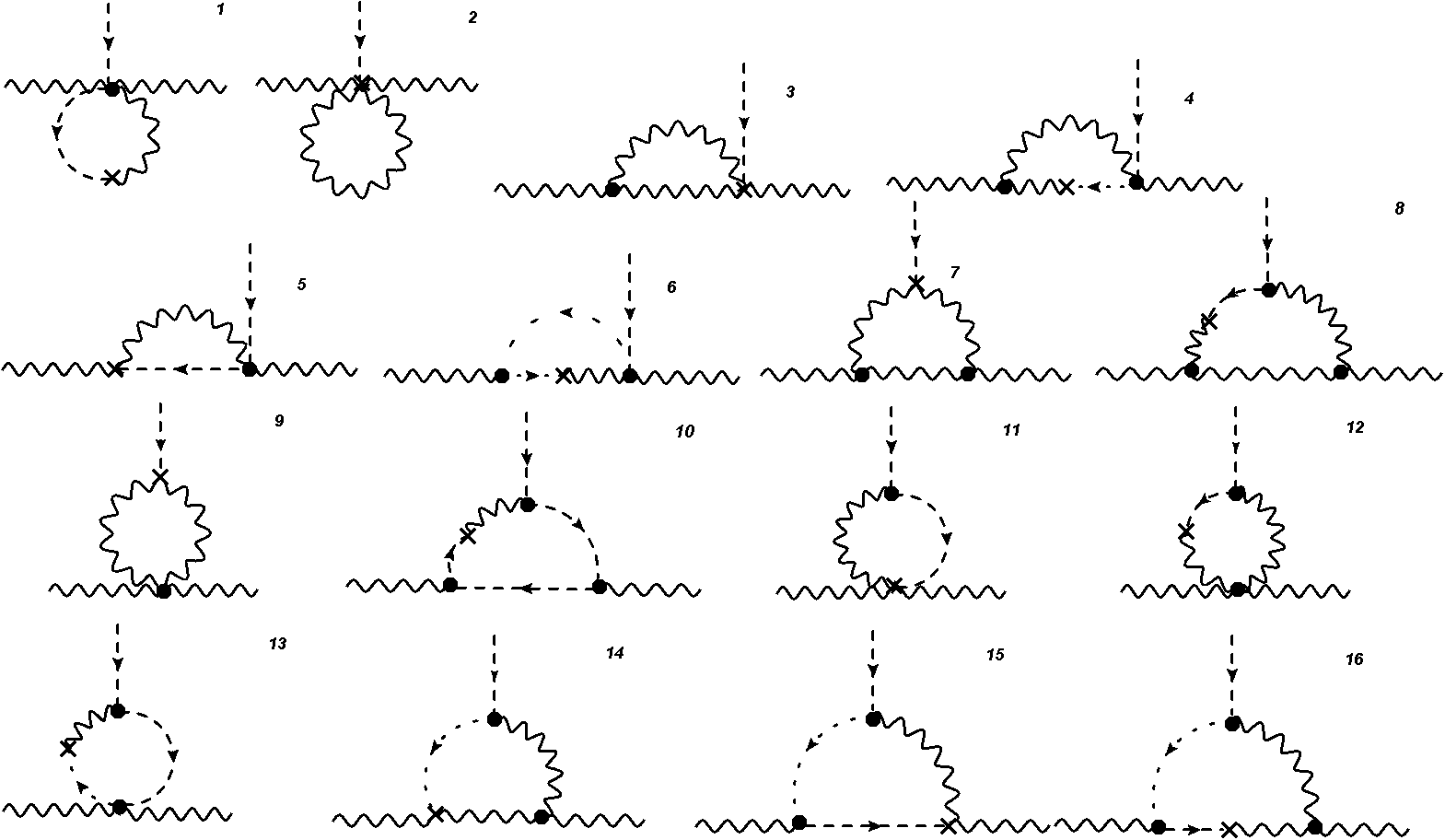}
\captionof{figure}{One-loop Feynman diagrams contributing to the three point Green's function of the supercurrent, $\langle u_\nu u_\rho S_\mu\bar \lambda  \rangle$\,.  A wavy (dashed) line represents gluons (gluinos). A cross denotes the insertion of the operator. Diagrams 1, 2, 3, 5, 6, 11, and 13 do not appear in dimensional regularization, but they contribute in the lattice regularization. Mirror versions of the diagrams must also be included.}
\label{fig3ptguu}
\end{center}

\section{Mixing matrix in Dimensional Regularization}

We use the $\MSbar$ renormalization scheme in order to calculate the elements of the mixing matrix, the first row of which renormalizes the supercurrent operator $S_\mu$. The divergent parts of the one-loop contributions are expected to contain tensorial structures of the tree-level Green's functions of the operators. Thus, in order to determine the renormalization and mixing coefficients we employ the expressions for the tree-level two- and three-point Green's functions of the operators $S_\mu$ and $T_\mu$ and all gauge-noninvariant operators which could mix with them, with an operator insertion at point $x$. The mixing matrix is a $14 \times 14$ block upper triangular matrix. The renormalized supercurrent can be written as a linear combination of these operators:
\begin{equation}
S_\mu^R  = Z_{S,S} S_\mu^B +z_{S,T} T_\mu^B + z_{S,A1} {\cal O}^B_{A1} + z_{S,B1} {\cal O}^B_{B1} + z_{S,B2} {\cal O}^B_{B2}  + \sum_{i=1}^{9} z_{S,Ci} {\cal O}^B_{Ci} 
\label{Zz} 
\end{equation}
Eq.~(\ref{Zz}) defines the first row of the mixing matrix. A similar expression holds for $T_\mu^R$. The renormalization function $Z = \openone + {\cal O}(g^2)\,$ and the mixing coefficients $z = {\cal O}(g^2)$ should more properly be denoted as $Z^{X,Y}$and $z^{X,Y}$, where $X$ is the regularization and $Y$ the renormalization scheme. Superscript $B$ stands for bare and $R$ for renormalized quantities. We are interested in calculating the one-loop renormalization function of $S_\mu$ and the mixing coefficients relevant to other operators with equal/lower dimensionality. In order to calculate the one-loop renormalization function and the mixing coefficients, we compute the amputated two-point Green's function of ${S}_{\mu}$ with one external gluino and one external gluon fields $\langle u_\nu \,{S}_{\mu}\, \bar\lambda \rangle _{amp}$, as well as three-point Green's functions with external gluino/ghost/antighost fields $\langle c \,{S}_\mu\,\bar c\, \bar\lambda \rangle _{amp}$ and with external gluino/gluon/gluon fields $\langle u_\nu u_\rho \,{S}_{\mu}\, \bar\lambda \rangle _{amp}$ (Fig.~\ref{fig3ptguu}).

We need also the renormalization functions of the gluon, gluino, ghost and coupling constant:
\begin{equation}
u_{\mu}^R = \sqrt{Z_u}\,u^B_{\mu}\,\,\,,\,\,\,
\lambda^R = \sqrt{Z_\lambda}\,\lambda^B\,\,\,,\,\,\,
c^R = \sqrt{Z_c}\,c^B\,\,\,,\,\,\,
g^R = Z_g\,\mu^{-\epsilon}\,g^B
\end{equation}
where $\mu$ is an arbitrary scale with dimensions of inverse length. For one-loop calculations, the distinction between $g^R$ and $\mu^{-\epsilon}\,g^B$ is inessential in many cases; we will denote both by $g$ in those cases. Our results are presented as functions of the $\MSbar$ scale $\bar\mu$ which is related to $\mu$ through: $\mu = \bar \mu \sqrt{e^{\gamma_E}/ 4\pi}$ \, ($\gamma_E = 0.57721\ldots$ is Euler's constant).

The renormalization conditions involve the renormalization factors of the external fields as well as parameters that show up in the bare GFs. The condition for the gluino-gluon GF of the operator ${S}_{\mu}$ is:
\begin{eqnarray}
\langle u_\nu^R \,{S}^R_{\mu}\, \bar\lambda^R \rangle _{amp} &=& Z_\lambda^{-1/2} \,Z_u^{-1/2} Z_{S,S} \langle u_\nu^B \,{S}^B_{\mu}\,\bar\lambda^B \rangle _{amp}+ z_{S,T}  \langle u_\nu^B \, {T_\mu}^B\, \bar\lambda^B \rangle _{amp}^{tree} \nonumber\\
&+& z_{S,A1} \langle  u_\nu^B \, {\cal O}^B_{A1} \, \bar\lambda^B \rangle _{amp}^{tree}  
+ z_{S,B1} \langle  u_\nu^B \, {\cal O}^B_{B1} \, \bar\lambda^B \rangle _{amp}^{tree} \nonumber\\
&+& z_{S,B2} \langle  u_\nu^B \, {\cal O}^B_{B2} \, \bar\lambda^B \rangle _{amp}^{tree} 
+ \sum_{i=1}^{6} z_{S,Ci} \langle u_\nu^B \,{\cal O}^B_{Ci} \, \bar\lambda^B \rangle _{amp}^{tree} + {\cal O}(g^4)
\label{2ptGFexprS}
\end{eqnarray}
A similar condition holds for the operator ${T}_{\mu}$. 
Results for the renormalization of the external fields and of the coupling constant have been already calculated in Ref.~\cite{CHPP:GluinoGlue}.

Imposing condition~(\ref{2ptGFexprS}) on the two-point functions is sufficient in order to obtain the renormalization of the supercurrent $Z_{S,S}$. The pole part in the continuum results of the one-loop two-point GF of $S_\mu$ (choosing zero momentum of the external gluino) is proportional to the tree-level GF of $S_\mu$ and thus there is no mixing with $T_\mu$, ${\cal O}_{A1}$, ${\cal O}_{C4}$, ${\cal O}_{C5}$:  $z_{S,T} = z_{S,A1} = z_{S,C4} = z_{S,C5} = 0$. Operators ${\cal O}_{C1}$ and ${\cal O}_{C2}$ are of lower dimensionality and they will not mix in the continuum regularization: $z_{S,C1} = z_{S,C2} = 0$. Demanding that the left-hand side of Eq.~(\ref{2ptGFexprS}) be finite leads to the determination of $Z_{S,S}^{DR,\MSbar}$.

The $\MSbar$-renormalized GFs stemming from the calculation of all the above bare GFs are essential ingredients in order to extract the lattice renormalization factors and mixing coefficients. 

\section{Mixing matrix on the lattice}

We make use of the Wilson formulation on the lattice, with the addition of the clover (SW) term for gluino fields. The Euclidean action ${\cal S}^{L}_{\rm SYM}$ on the lattice becomes:
\begin{eqnarray}
{\cal S}^{L}_{\rm SYM}=a^{4}\sum_{x}&\bigg[&\frac{N_{c}}{g^{2}}\sum_{\mu, \nu}\bigg(1-\frac{1}{N_{c}}TrU_{\mu \nu}\bigg) + \sum_{\mu}\bigg(Tr\bigg(\bar\lambda\gamma_{\mu} D_{\mu} \lambda\bigg) -\frac{a r}{2}Tr\bigg(\bar\lambda D^{2}\lambda \bigg)\bigg) \nonumber\\
 &-& \sum_{\mu, \nu}\bigg(\frac{c_{\rm SW} \ a}{4}\bar\lambda^{\alpha}\sigma_{\mu \nu}\hat{F}_{\mu \nu}^{\alpha \beta}\lambda^{\beta}\bigg) + m_0 Tr \bigg(\bar\lambda \lambda\bigg)\bigg]
\label{susylagrLattice}
\end{eqnarray}
The field strength tensor $\hat{F}_{\mu \nu}^{\alpha \beta}$ in the adjoint representation and the covariant derivatives are defined in a standard way (see, e.g.,  Ref.~\cite{CHPP:GluinoGlue}).

The ``Lagrangian mass'', $m_0$, is a free parameter in principle and represents the bare gluino mass. This term breaks supersymmetry softly. All renormalization functions which we will be calculating, must be evaluated at vanishing renormalized mass, that is, when $m_0$ is set equal to the critical value ensuring a massless gluino in the continuum limit. However, since our calculations are at one loop order this critical value is irrelevant, being already of order $g^2$. Just as in the continuum, a gauge-fixing term, together with the compensating ghost field term, must also be added to the action; these terms are the same as in the non-supersymmetric case~\cite{Kawai:1980ja}. Similarly, a standard ``measure'' term must be added to the action, in order to account for the Jacobian in the change of integration variables: $U_\mu \to u_\mu$\,. Further details of the lattice action can be found in Ref.~\cite{Costa:2017rht}. For the lattice discretization of operators $S_\mu$ and $T_\mu$ we employ a standard clover version of the gauge field strength $u_{\mu \nu}$ in the fundamental representation.  

We have computed to 1 loop all 2- and 3-pt lattice Green's functions mentioned above, thus providing an evaluation of the $\MSbar$ renormalization functions and mixing coefficients, independently of the method presented below. The renormalization functions $Z^{L,\MSbar}$ can be readily extracted from these functions. These results, along with comparisons with the non-perturbative determinations described below, will be presented in a longer write-up of this work.

\section{GIRS scheme}

In order to extract non-perturbative physical results from numerical simulations on the lattice, we employ a non-perturbative gauge-invariant renormalization scheme (GIRS)~\cite{Costa:2021iyv}, which is applicable in both continuum and lattice regularizations so as to make contact with the continuum schemes. In particular, we consider Green's functions which involve products of two gauge-invariant operators, ${\cal O}_1$\,, ${\cal O}_2$\,, at distinct spacetime points, in such a way as to avoid potential contact terms:
\begin{equation}
\langle {\cal O}_1 (x) \ {\cal O}_2 (y) \rangle, \qquad (x \neq y).
\label{GIRScons}
\end{equation}
Only the mixing of gauge-invariant operators is relevant in this case, resulting in a $2 \times 2$ mixing matrix, which relates the bare to the renormalized operators $S_{\mu}$ and $T_{\mu}$\,:
\begin{gather}
 \begin{pmatrix} {S}^R_\mu & \\[2ex] {T}^R_{\mu} \end{pmatrix}
 =
\begin{pmatrix}
   Z_{SS} &
   Z_{ST} \\[2ex]
   Z_{TS} &
   Z_{TT} 
   \end{pmatrix}
   \begin{pmatrix} {S}^B_{\mu} & \\[2ex] {T}^B_{\mu} \end{pmatrix}.
   \label{mixing matrix}
\end{gather}
In order to determine the 4 elements of the mixing matrix $Z$ we need 4 conditions.
\begin{itemize}
\item Three conditions can be imposed by considering expectation values between the two mixing operators:
\begin{equation}
G^{S\,S}_{\mu \nu}(x, y) \equiv \langle S_\mu (x) \ \overline{ S}_\nu (y) \rangle\,\,\,,\,\,\,
G^{T\,T}_{\mu \nu}(x, y) \equiv \langle T_\mu (x) \ \overline{ T}_\nu (y) \rangle\,\,\,,\,\,\,
G^{S\,T}_{\mu \nu}(x, y) \equiv \langle S_\mu (x) \ \overline{ T}_\nu (y) \rangle.
\label{GFs}
\end{equation}
In explicit form, the bar (charge conjugate) operators are:
\begin{equation}
\overline{ S}_\mu \equiv \,{\rm tr}_c (\, \bar \lambda u_{\nu\,\rho}) \gamma_\mu \sigma_{\nu \rho}\,\,\,,\,\,\,
\overline{ T}_\mu \equiv 2\, {\rm tr}_c(\,\bar \lambda u_{\mu\,\nu})\gamma_\nu.
\label{supercurrent operators}
\end{equation}
\item A fourth condition can be obtained by considering two-point Green's functions involving products of $S_\mu$ (or $T_\mu$) with other gauge-invariant operators, such as the Gluino-Glue operator $\mathcal{O}$, e.g.:
\begin{equation}
G^{{\cal O}\,S}_{\mu}(x, y) \equiv \langle {\cal O} (x) \ \overline{S}_\mu (y) \rangle,
\label{GFGgS}
\end{equation}
where ${\cal O} \equiv \sigma_{\mu \nu} \,{\rm{tr}}_c (\,u_{\mu \nu} \lambda)$.
\item A condition for calculating the renormalization factor of ${\cal O}$, can be obtained by considering the two-point Green's function involving the product of two Gluino-Glue operators.
\end{itemize}

Below, we present our one-loop results for the $\MSbar$-renormalized Green's functions in dimensional renormalization. We note that the evaluation of these Green's functions to order $g^{2n}$ involves diagrams with $(n+1)$ loops. This is a non-negligible price to pay. However, all these diagrams involve massless fields and (upon expressing them in momentum space) only one incoming/outgoing momentum and thus can be evaluated to very high perturbative order.
\begin{eqnarray}
{\left[G^{S\,S}_{\mu \nu}(x, y)\right]}^{\MSbar} &=& \frac{2 (N_c^2 -1)}{3 \pi^4 {(z^2)}^4} (3 - 5 \frac{g^2_{\MSbar}}{16 \pi^2} N_c) (4 s_{\mu \nu}^{[3]} + s_{\mu \nu}^{[4]}), \label{GSSGIRSGF} \\
{\left[G^{T\,T}_{\mu \nu}(x, y)\right]}^{\MSbar} &=& \frac{(N_c^2 -1)}{6 \pi^4 {(z^2)}^4} \Big[(3 - 5 \frac{g^2_{\MSbar}}{16 \pi^2} N_c) 2 (2 s_{\mu \nu}^{[3]} - s_{\mu \nu}^{[4]}) \nonumber\\
  && \hspace{0.3cm} + \,9\,\frac{g^2_{\MSbar}}{16 \pi^2} N_c \Big(2 s_{\mu \nu}^{[1]} +2 s_{\mu \nu}^{[2]} - 3 (3 + 4 \gamma_E - 4 \ln (2) + 2 \ln (\bar{\mu}^2 z^2)) s_{\mu \nu}^{[4]}\Big)  \Big], \qquad \\
{\left[G^{S\,T}_{\mu \nu}(x, y)\right]}^{\MSbar} &=& \frac{(N_c^2 -1)}{3 \pi^4 {(z^2)}^4} \Big[(3 - 5 \frac{g^2_{\MSbar}}{16 \pi^2} N_c) (4 s_{\mu \nu}^{[3]} + s_{\mu \nu}^{[4]}) + 18 \frac{g^2_{\MSbar}}{16 \pi^2} N_c (s_{\mu \nu}^{[2]} - s_{\mu \nu}^{[4]})\Big], \qquad \\
{\left[G^{{\cal O}\,S}_{\mu}(x, y)\right]}^{\MSbar} &=& \frac{(N_c^2 -1)}{\pi^4 {(z^2)}^4} \,12 \,\frac{g^2_{\MSbar}}{16 \pi^2} \,N_c \,\sigma_{\mu \rho}\,\, z_\rho,
\end{eqnarray}
where $z \equiv y - x$ and we have used the following notation:
\begin{equation}
s_{\mu \nu}^{[1]} (z) \equiv \gamma_\mu z_\nu\,\,\,,\,\,\,
s_{\mu \nu}^{[2]} (z) \equiv \gamma_\nu z_\mu\,\,\,,\,\,\,
s_{\mu \nu}^{[3]} (z) \equiv (\delta_{\mu \nu} - 2 \frac{z_\mu z_\nu}{z^2}) \zslash\,\,\,,\,\,\,
s_{\mu \nu}^{[4]} (z) \equiv \gamma_\mu \zslash \gamma_\nu
\label{GFsNotation}
\end{equation}
Note that in the presence of mass, the above Green's functions may also contain the structures of Eq.(\ref{GFsNotation}) multiplied by an extra $\zslash$. 

There is a variety of ways to impose renormalization conditions in GIRS. In what follows, we consider the following set of conditions, in which we integrate over the spatial components of $z = y - x = (\vec{z},t)$\,:

\begin{eqnarray}
\int d^3 \vec{z} \ {\rm Tr} \{ {\left[G^{S\,S}_{\mu \nu}(x, y)\right]}^{{\rm GIRS}} P_{\nu \mu} \} &=& \int d^3 \vec{z} \ {\rm Tr} \{ {\left[G^{S\,S}_{\mu \nu}(x, y)\right]}^{\rm tree} P_{\nu \mu} \}, \label{GIRS2_cond1} \\
\int d^3 \vec{z} \ {\rm Tr} \{ {\left[G^{T\,T}_{\mu \nu}(x, y)\right]}^{{\rm GIRS}} P_{\nu \mu} \} &=& \int d^3 \vec{z} \ {\rm Tr} \{ {\left[G^{T\,T}_{\mu \nu}(x, y)\right]}^{\rm tree} P_{\nu \mu} \}, \label{GIRS2_cond2} \\
\int d^3 \vec{z} \ {\rm Tr} \{ {\left[G^{S\,T}_{\mu \nu}(x, y)\right]}^{{\rm GIRS}} P_{\nu \mu} \} &=& \int d^3 \vec{z} \ {\rm Tr} \{ {\left[G^{S\,T}_{\mu \nu}(x, y)\right]}^{\rm tree} P_{\nu \mu} \}, \label{GIRS2_cond3} \\
\int d^3 \vec{z} \ {\rm Tr} \{ {\left[G^{S\,{\cal O}}_{\mu}(x, y)\right]}^{{\rm GIRS}} P_{\mu} \} &=& \int d^3 \vec{z} \ {\rm Tr} \{ {\left[G^{S\,{\cal O}}_{\mu}(x, y)\right]}^{\rm tree} P_{\mu} \}.
\label{GIRS2_cond4}
\end{eqnarray}
where $P_{\nu \mu}$, $P_{\mu}$ are projectors acting on the Dirac space; the repeated indices $\mu, \nu$ are not summed over. There are various options for the choice of indices, some of which may be better than others from the simulation point of view. In particular,

\begin{itemize}
\item  in Eq.~(\ref{GIRS2_cond4}) there are 2 options: $\mu = t$ or $s$ [where t (s) denotes temporal (spatial) direction],
\item  in Eqs. (\ref{GIRS2_cond1} -- \ref{GIRS2_cond3}) there are a priori 5 options for each condition: $\mu = \nu = t$, $\mu = \nu = s$, $(\mu = t, \nu = s)$, $(\mu = s, \nu = t)$, $(\mu = s, \nu = s')$ [where $t, s, s'$ are temporal and two different spatial directions, respectively].
\end{itemize}
In all the above choices, the projectors which can lead to a solvable system of conditions, are:
\begin{equation}
P_{\nu \mu} = \gamma_\nu \gamma_4 \gamma_\mu\qquad, \qquad P_{\mu} = \gamma_4 \gamma_\mu
\label{proj}
\end{equation}

The conversion factors between the $\MSbar$ and GIRS scheme follow immediately from Eqs.(\ref{GSSGIRSGF} -- \ref{GIRS2_cond4}).

As we are interested in applying GIRS in lattice simulations, the scale $t$ may be chosen to satisfy the condition $a \ll |t| \ll \Lambda_{\rm SYM}^{-1}$, where $a$ is the lattice spacing and $\Lambda_{\rm SYM}$ is the SYM physical scale; this condition guarantees that discretization effects will be under control and simultaneously we will be able to make contact with (continuum) perturbation theory.

\section{Non-perturbative results}
As a first test of the non-perturbative determination, we have considered SU(2) $\mathcal{N}=1$ SYM with the lattice action \eqref{susylagrLattice}. 
The gluinos are described by Wilson fermions ($r=1$) in the adjoint representation. The Wilson-Dirac operator of Eq.~\eqref{susylagrLattice} takes hence the following form,
\begin{equation*}
            D_w = 1- \kappa \big[(1-\gamma_\mu)(V_\mu(x))\delta_{x+\mu,y} + (1+\gamma_\mu)(V^ \dagger{}_\mu(x+\mu))\delta_{x-\mu,y}\big]\, ,
\end{equation*}
where the gauge links are in the adjoint representation. The hopping parameter $\kappa$ is related to the bare gluino mass $m_0$ by $\kappa =1/(2m_0+8)$.

The ensemble of gauge configurations has been generated in earlier studies and presented first in \cite{Bergner:2015adz}. 
The lattice size is $N_s=24$ in spatial and $N_t=48$ in temporal direction. For a further reduction of lattice artefacts, a tree-level Symanzik improved plaquette action and stout smeared links in the Dirac operator are used. The mass parameter and the inverse gauge coupling are $\kappa=0.14925$ and $\beta=1.75$. Further details can be found in \cite{Bergner:2015adz}. The sign problem appearing for simulations with Majorana fermions, see \cite{Bergner:2011sap}, is not relevant for the considered parameters.

We have computed the bare correlation functions between the supercurrent operators \eqref{supercurrent operators} for time-like separations ($\mu=4$) and the gluino-glue operator defined with only spatial indices  ${\cal O}\equiv\sigma_{ij} \,{\tr}_c (\,u_{ij} \lambda)$.

\begin{figure}[h!]
    \centering
    \begin{subfigure}[b]{0.45\textwidth}
        \centering
        \includegraphics[width=\textwidth]{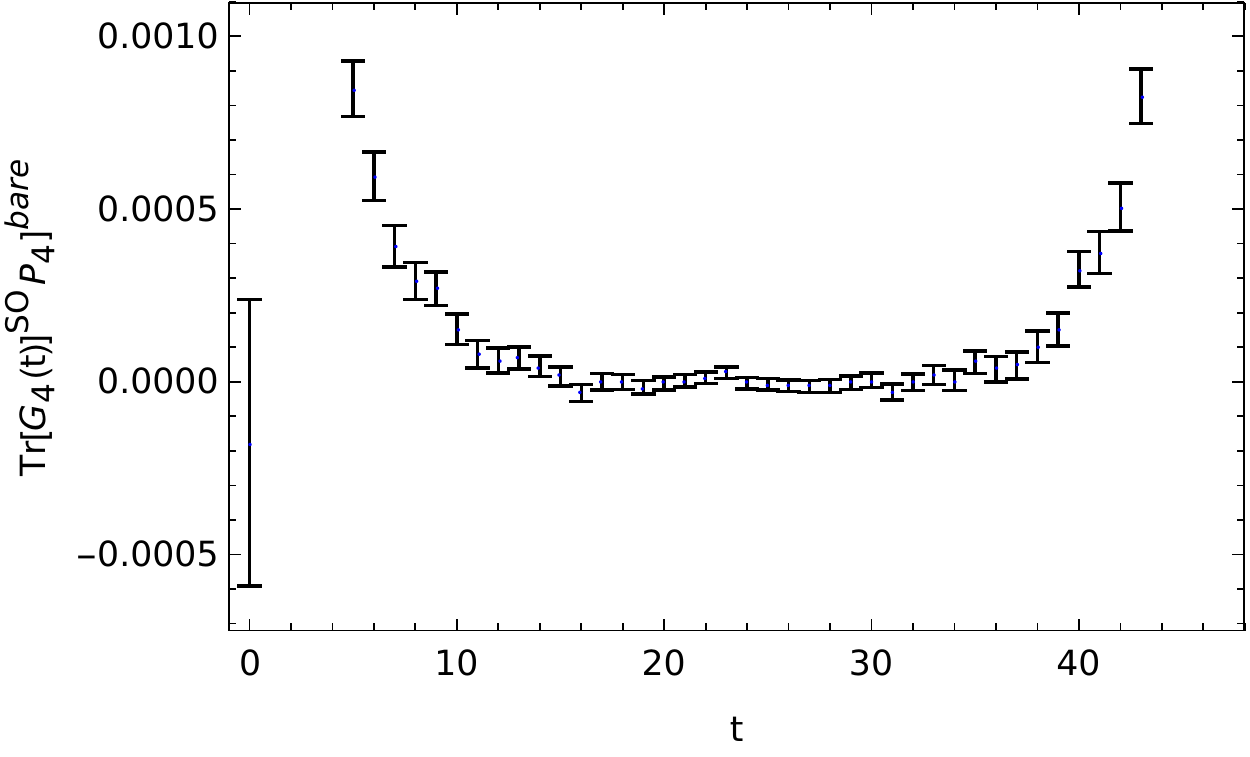}
    \end{subfigure}
    \hfill
    \begin{subfigure}[b]{0.45\textwidth}
        \centering  
        \includegraphics[width=0.99\textwidth]{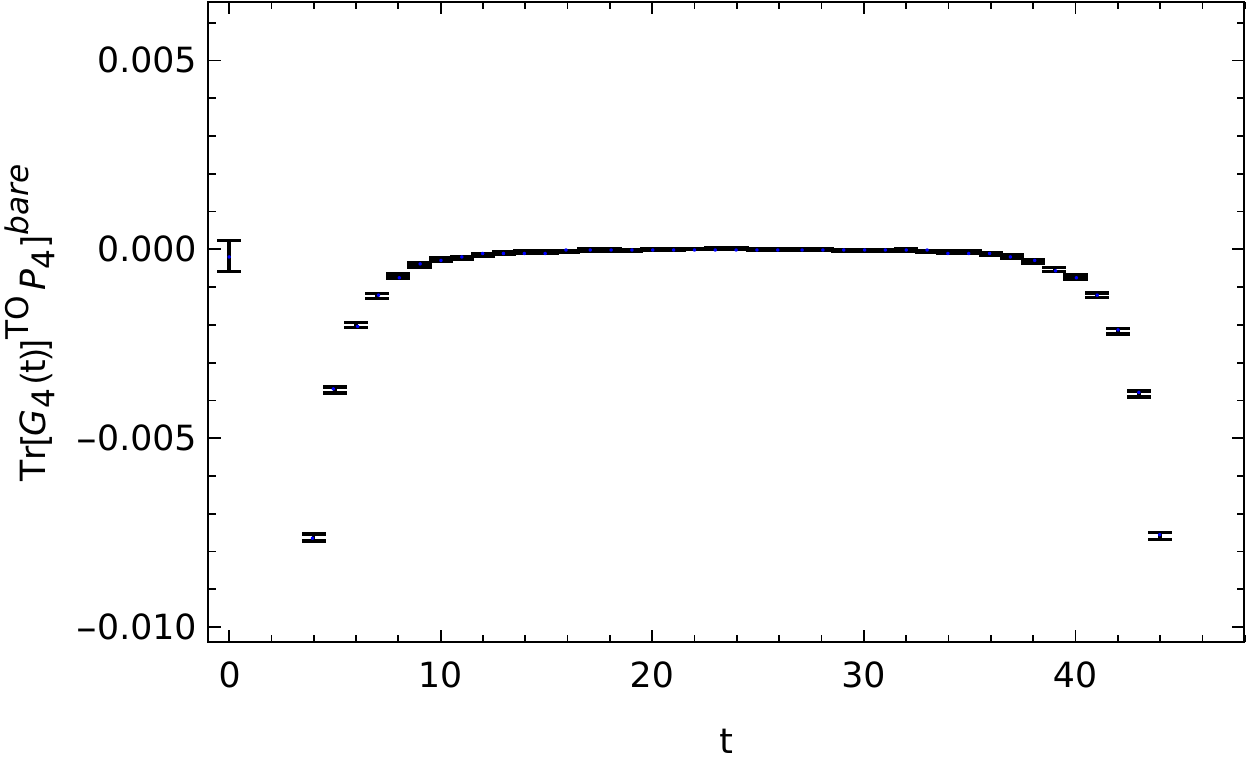}
    \end{subfigure}
\caption{Correlation functions integrated over the spatial component as appearing in Eqs.~(\ref{GIRS2_cond1} -- \ref{GIRS2_cond4}), including the  operator ${\cal O}\equiv\sigma_{ij} \,{\rm{tr}}_c (\,u_{ij} \lambda)$  on a $24x48$ lattice.}
\label{fig: GTO and GSO}
\end{figure}

\begin{figure}[h!]
        \centering
        \includegraphics[width=12cm]{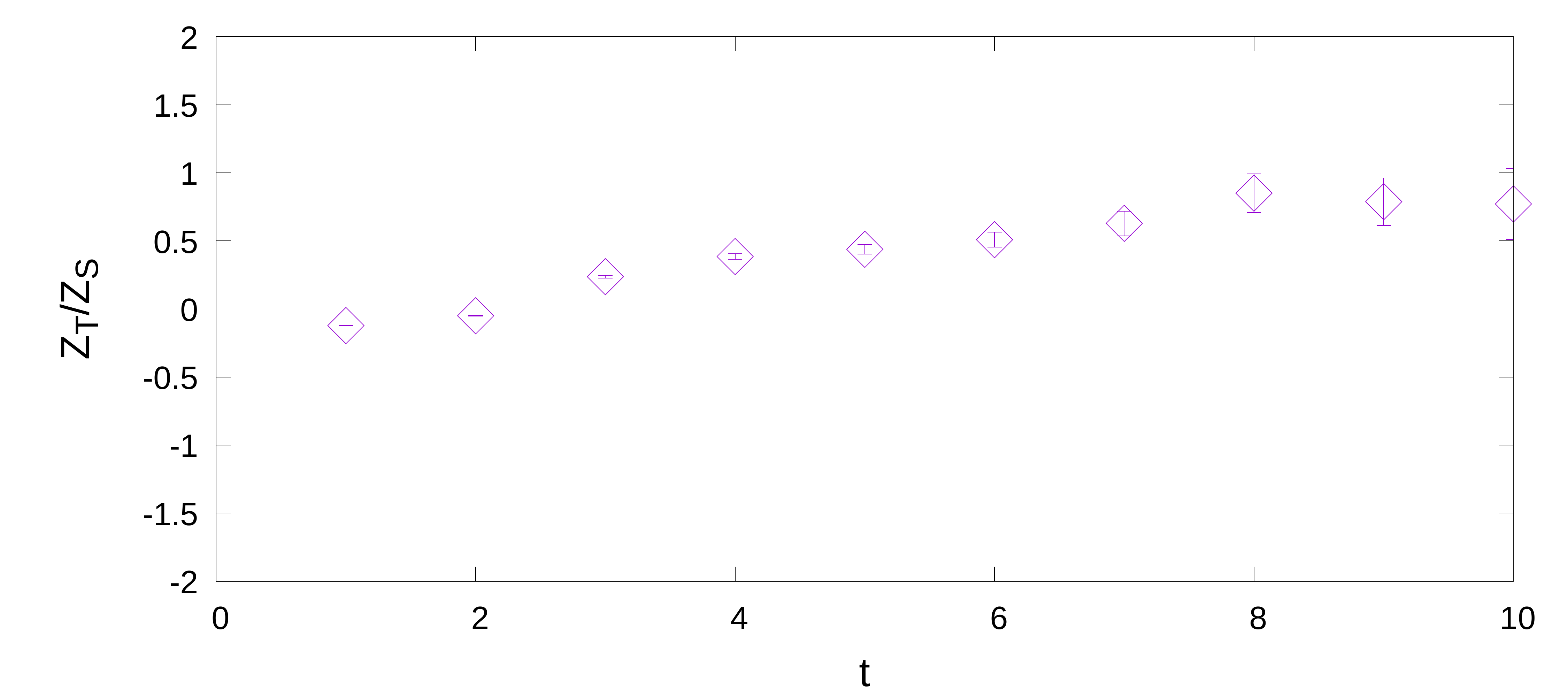}
        \caption{ $Z_T/Z_S$ as a function of time separation $t$ obtained through a Jackknife analysis. The relevant data values of the plot lie on the interval $t\in[4,6]$, where $t$ is big enough such that contact terms are suppressed and small enough such that the statistical fluctuations are not strongly overcoming the signal.}
        \label{Zt_zs}
\end{figure}

We have applied the GIRS conditions Eqs.~(\ref{GIRS2_cond1} -- \ref{GIRS2_cond4}) to obtain the renormalized correlation functions from the lattice data. For a first preliminary result, we just focused on the renormalization properties of the $S$ and $O$ operator
\begin{equation*}
    S_\mu^{ren} = Z_S S_\mu^{bare} + Z_T T_\mu^{bare} \quad , \quad {\cal O}^{ren} = Z_{\cal O} {\cal O}^{bare},
\end{equation*}
which in Eq.~\eqref{GIRS2_cond4}

\begin{eqnarray*}
\int d^3 \vec{z} Z_{\cal O}(Z_{S}\text{Tr}[G_\mu^{S{\cal O}}(z)P_\mu]^{bare} + Z_{T}\text{Tr}[G_\mu^{T{\cal O}}(z)P_\mu]^{bare})=\int d^3 \vec{z} \text{Tr}[G^{S{\cal O}}_\mu (z) P_\mu]^{tree}=0.
\end{eqnarray*}
From here we see that the $Z_{\cal O}$ renormalization factor drops out and, using only the $G_{S{\cal O}}$ and $G_{T{\cal O}}$ data, we can already solve for the $Z_T/Z_S$ quotient at each time slice $t$ (Fig.~\ref{Zt_zs}). We observe a plateau-like behaviour of $Z_T/Z_S$ in the interval $t\in [4,6]$. This is in accordance with the fact that the final renormalization factors should be independent of the time separation $t$.

As already mentioned, for this first numerical test we used the spatial operator ${\cal O}\equiv\sigma_{ij} \,{\tr}_c (\,u_{ij} \lambda)$, however also time components are needed in order to apply the GIRS condition Eq.~\eqref{GIRS2_cond4}. This complete analysis is left for the further final studies. 

Remarkably, the signal to noise ratio at the considered time separations is relatively low and further improvements are expected with a full statistics and applying the complete GIRS renormalization conditions Eqs.~(\ref{GIRS2_cond1} -- \ref{GIRS2_cond4}). With the current preliminary data, we have not done a full  conversion to $\MSbar$ scheme. However, the current data is already promising and we are currently working towards a full non-perturbative determination of the renormalization factors.

{\bf Acknowledgements:} M.C., H.P. and A.S. acknowledge financial support from the Cyprus Research and Innovation Foundation (RIF) under contract number EXCELLENCE/0918/0066. G.B. and I.S. acknowledge support from the Deutsche Forschungsgemeinschaft (DFG) Grant No.~BE 5942/3-1 and 5942/4-1.

\end{document}